\documentclass[sigconf]{acmart}
\usepackage{todonotes}

\AtBeginDocument{%
  }

\setcopyright{acmlicensed}
\copyrightyear{2025}
\acmYear{2025}
\acmDOI{XXXXXXX.XXXXXXX}

\acmConference[Conference acronym 'XX]{Make sure to enter the correct
  conference title from your rights confirmation emai}{June 03--05,
  2018}{Woodstock, NY}
\acmISBN{978-1-4503-XXXX-X/18/06}

\DeclareMathOperator*{\argmax}{argmax}

\newtheorem{definition}{Definition}
\begin{document}

\title{Achieving Socio-Economic Parity through the Lens of EU AI Act}

\author[1]{Arjun Roy}
\authornote{Both authors contributed equally to this research.}
\email{arjun.roy@unibw.de}
\orcid{0000-0002-4279-9442}
\affiliation{%
\institution{Dept. Mathematics \& CSc., Freie Universität Berlin}
  \institution{RI CODE, Universität der Bundeswehr München}
  \country{Germany}
}
\author{Stavroula Rizou}
\authornotemark[1]
\email{rizstavroula@iti.gr}
\affiliation{%
  \institution{Information Technologies Institute, CERTH}
  \country{Greece}
}

\author{Symeon Papadopoulos}
\email{papadop@iti.gr}
\affiliation{%
  \institution{Information Technologies Institute, CERTH}
  \country{Greece}
}
\author{Eirini Ntoutsi}
\email{eirini.ntoutsi@unibw.de}
\affiliation{%
  \institution{RI CODE, Universität der Bundeswehr München}
  \country{Germany}
}




\renewcommand{\shortauthors}{Roy et al.}

\begin{abstract}
Unfair treatment and discrimination are critical ethical concerns in AI systems, particularly as their adoption expands across diverse domains. Addressing these challenges, the recent introduction of the EU AI Act establishes a unified legal framework to ensure legal certainty for AI innovation and investment while safeguarding public interests,  such as health, safety, fundamental rights, democracy, and the rule of law (Recital 8). 
The Act encourages stakeholders to initiate dialogue based on existing AI fairness notions to address discriminatory outcomes of AI systems. However, these fairness notions often overlook the critical role of Socio-Economic Status (SES) and fail to capture the compounded effects of economic privilege, inadvertently perpetuating biases that favour the economically advantaged. This oversight is particularly concerning given that the principles of equalization advocate for equalizing resources or opportunities to mitigate disadvantages beyond an individual's control. 
While provisions for discrimination are laid down in the AI Act, specialized directions should be broadened, particularly in addressing economic disparities perpetuated by AI systems. 
In this work, we explore the limitations of popular AI fairness notions using a real-world dataset (Adult), highlighting their limitations, particularly their inability to address SES-driven disparities. To address this gap, we propose a novel fairness notion, Socio-Economic Parity (SEP), which incorporates SES and promotes positive actions for underprivileged groups while accounting for factors within individual's control, such as working hours, which can serve as a proxy for effort. We define a corresponding fairness measure based on this notion, and optimize a model constrained by SEP to demonstrate its practical utility. Our empirical results demonstrate the effectiveness of our approach in mitigating SES-driven biases.  
By analyzing the AI Act with our method, we lay a foundation for aligning AI fairness with SES factors while ensuring legal compliance in AI decisions.

\end{abstract}



\maketitle

\section{INTRODUCTION}
\begin{quote}
    ``\textit{We seem to be back in the Dickensian world of Hard Times, with the haves facing off against the increasingly alienated have-nots, with no resolution in sight}''~\cite{banerjee2019good}
\end{quote}
Unfair treatment and discrimination \cite{ai2019high, information_commissioners_office_ico_guidance_2024} in AI systems — often described as AI bias or unfairness — are widely recognized as significant ethical concerns \cite{access_now_human_2018, latonero_governing_2018, muller_impact_2020}. The pervasive integration of AI across various societal and economic domains has amplified the urgency of addressing these issues. This is further underscored by the increasing global adoption of AI-related legal frameworks \cite{etchemendy_ai_2024}, including the landmark EU AI Act proposed in 2021, which catalyzed a 1,200\% increase\footnote{Based on public data for 2021 and 2022 from the AI Index 2024 Annual Report by Stanford University \cite{ai_index_steering_committee_institute_for_human-centered_ai_stanford_university_ai_nodate}.} in regulatory activities in the following year. While the Act provides a comprehensive framework for addressing discrimination, its scope should be broadened to include specialized directions targeting economic disparities perpetuated by AI systems~\cite{ganty2021expanding}. These disparities are deeply rooted in Socio-Economic Status (SES), which disproportionately affects historically marginalized groups~\cite{ganty2021expanding}. For instance, individuals subjected to discrimination based on traditional grounds, such as sex and race, are often overrepresented among economically disadvantaged populations~\cite{ganty2021poverty}. This intersectionality of SES with existing vulnerabilities further exacerbates systemic inequities~\cite{atrey2018intersectional}, violating principles of equalization~\cite{dworkin2018equality}, which advocate for equalizing resources or opportunities to mitigate disadvantages beyond an individual's control. Notably, at the EU level, the pay gap between men and women was 13\% in 2020 and has decreased by only 2.8 percentage points over a decade~\cite{EqualPayFactsheet2022}. 

Despite advancements in AI fairness metrics \cite{barocas_fairmlbook, castelnovo2022clarification}, most fail to capture the compounded effects of socio-economic privilege. This limitation creates critical gaps in addressing SES-driven disparities that produce unfair advantages across protected attributes such as sex~\cite{kilbourne1997ses_gender} or race~\cite{williams2017wealth,kochhar2014wealth}. To bridge this gap, we propose a novel fairness notion, Socio-Economic Parity (SEP), along with its stricter variant, Conditional Socio-Economic Parity (CSEP), explicitly incorporating SES into fairness considerations. Our contributions include introducing SEP and CSEP, demonstrating their effectiveness in mitigating SES-driven bias through evaluations on the Adult dataset~\cite{adult}, and analyzing their alignment with the AI Act to bridge fairness-aware ML research with regulatory frameworks.  
Using this approach, we emphasize positive actions for underprivileged groups while accounting for individual-level factors~\cite{cappelen2009rewarding} (e.g., working hours).  
We demonstrate that existing fairness notions—such as Equal Opportunity~\cite{hardt2016equality}, Demographic Parity~\cite{dwork2012fairness}, and Conditional Demographic Parity (CDP)~\cite{toon2012book}—fail to address SES-driven disparities and inadvertently reinforce biases favouring economically privileged groups. Our empirical results highlight the effectiveness of SEP and CSEP in mitigating SES-driven bias. Furthermore, through an analytical study of the AI Act in parallel with our proposed methods, we establish a foundation for fostering dialogues that align AI fairness in socio-economic applications, 
emphasizing the importance of addressing SES-driven disparities. 

The remainder of the paper is organized as follows: We start by presenting the investigation over the protected grounds in the context of SES under EU law (Sec. \ref{sec.def_grounds}). Sec.~\ref{sec.background} contextualizes socio-economic disparities through the existing fairness definitions, underlying challenges, and the driving motivations and objectives behind addressing them. Sec.~\ref{sec.method} outlines  
the proposed solution, focusing on key variables, and thresholds for fairness tests. Sec.~\ref{sec.features that define} examines the proposed notion through justifications, drawing on the AI Act and outlining also the legal framework of anti-discrimination law, which the regulation assigns with addressing specific issues related to AI bias. Sec.~\ref{sec.standartization} discusses the implementation of the assessments of the AI Act through fairness measures. Sec.~\ref{sec.conclude} concludes the paper with challenges and limitations.



\section{PROTECTED, NON-PROTECTED GROUNDS AND PROXIES IN THE CONTEXT OF SES UNDER EU ANTI-DISCRIMINATION LAW}\label{sec.def_grounds}
The investigation of protected grounds\footnote{Anti-discrimination law provides protection on the grounds of protected attributes \cite{alvarez2024policy}.}, which can affect the identification of various vulnerable groups under SES, according to EU non-discrimination law, is based on Article 2 of the Treaty on European Union (TEU) and Articles 10 and 19\footnote{The list of Article 19 includes the following protected grounds: sex, racial or ethnic origin, religion or belief, disability, age or sexual orientation.} of the Treaty on the Functioning of the European Union (TFEU), while EU secondary law prohibits discrimination in relevant adopted Directives\footnote{Directive 2000/43/EC against discrimination on grounds of race and ethnic origin; Directive 2000/78/EC against discrimination at work on grounds of religion or belief, disability, age or sexual orientation; Directive 2006/54/EC equal treatment for men and women in the field of employment and occupation; Directive 2004/113/EC equal treatment for men and women in the access to and supply of goods and services.} on defined grounds \cite{xenidis2020tuning}. Additionally, Article 21 of the EU Charter of Fundamental Rights provides a broader, non-exhaustive list\footnote{The list of Article 21 includes the following protected grounds:“…\textit{on any ground such as sex, race, colour, ethnic or social origin, genetic features, language, religion or belief, political or any other opinion, membership of a national minority, property, birth, disability, age or sexual orientation}.”}. However, it should be mentioned that this extended list in the Charter is limited solely to discrimination by the institutions and bodies of the EU, when exercising powers conferred under the Treaties and by Member States only when they are implementing Union law \cite{noauthor_article_2015}, indicating its limited scope when also being examined in the context of algorithmic fairness. In parallel, the European Convention on Human Rights (ECHR) law of the Council of Europe also plays a key role on the protected grounds under Article 14\footnote{The list of Article 14 includes the following protected grounds:“…\textit{on any ground such as sex, race, colour, language, religion, political or other opinion, national or social origin, association with a national minority, property, birth or other status.}”} in the EU environment, as the Treaty of Lisbon permitted the EU to accede to the ECHR \cite{ECHR,douglas2011european}. The non-exhaustive list of Article 14 does not include SES as a protected ground either, although, it allows potential expansion by the European Court of Human Rights \cite{european2018handbook,tobler2014equality}. Nevertheless, it should be noted that the European Court of Human Rights receives claims from victims of rights violations by States Parties to the Convention, with only a potential indirect effect on private actors \cite{gerards2022protected}.

Therefore, it becomes clear that SES is not considered a protected ground under EU law. It is nevertheless protected in the national law of certain EU Member States \cite{xenidis2020tuning}. Namely, 10\footnote{Belgium (social origin and wealth/income/property), Bulgaria (social status), Croatia (property, social status, social origin), France (economic vulnerability), Greece (social status), Hungary (social origin, financial status, part-time nature of employment), Lithuania (social status), Romania (social status), Slovakia (social origin, property), Slovenia (social standing, economic situation).} of the 27 EU Member States, according to the 2019 comparative analysis of non-discrimination law in Europe \cite{doi/10.2838/797667}, incorporate grounds related to SES into their anti-discrimination legislation with relative terminology as follows: social status, property, economic vulnerability, social origin, social standing, economic situation, profession and part-time employment. Apart from the investigation of protected grounds in the EU legal environment, and particularly in the case of SES, it should be noted that there are also non-protected categories that can serve as proxies for a protected ground \cite{xenidis2020eu}. Algorithms effectively identify correlations between factors that may not be identical to the protected ground; nevertheless, when considered in conjunction \cite{gerards2022protected}, these correlations can highlight the significance and the interference of proxies with the protected grounds and thus justify the existence of discrimination.

\section{PROBLEM BACKGROUND AND MOTIVATION}\label{sec.background}
We consider a supervised fairness-aware machine learning setup where a dataset \(\mathcal{D}\) is drawn from an independent and identically distributed (i.i.d.) sample space \(\mathcal{P}(X, S, Y)\). Here, \(X\) denotes unprotected attributes (e.g., education, job experience), \(S\) represents protected attributes (e.g., race, sex), and \(Y\) is the target class (e.g., loan approval). Protected attributes in \(S\) identify historically discriminated groups \cite{fundamental_rights_2007}, which must be safeguarded in automated decision-making \(E(h(x))\).
\(\mathbb{E}(h(x))\) is the expected favourable outcome from a predictor \(h()\).
Without loss of generality, we assume \(Y \in \{0,1\}\), with \(Y=1\) indicating favourable outcomes. 
Using this setup, fairness notions can be generalized as:
\begin{equation}\label{eq:generic_fair}
\mathbb{E}(h(x)\mid C)=\mathbb{E}(h(x)\mid s,C)), ~~\forall (x,s,y)\in\mathcal{D} 
\end{equation}\normalsize
where \(s\) denotes protected identity (group or subgroup), and \(C\) defines the fairness condition~\cite{RoyFact23}. 
This formula ensures that models favourable outcomes are comparable across groups under the specified fairness condition.

While numerous studies have explored various fairness notions~\cite{alves2023survey_notions}, their evaluations~\cite{RoyFact23,manios2024towards}, and underlying concepts~\cite{jaime2024facctethnic}, to the best of our knowledge, none have specifically addressed the issue of socio-economic status (SES)-driven bias in the context of achieving demographic fairness. 
To understand the drawbacks of existing fairness definitions in recognizing SES disparities, we focus on three popular~\cite{alves2023survey_notions} notions: Equal Opportunity (EP)~\cite{hardt2016equality}, Demographic Parity (DP)~\cite{dwork2012fairness}, and Conditional Demographic Parity (CDP)~\cite{kamiran2013quantifying}. EP ensures fairness in true positive rates across groups and aligns with correctness in predictions. DP eliminates outcome dependency on protected attributes, CDP extends DP by conditioning equality on additional unprotected attributes like occupation, tackling nuanced disparities like those in ``redlining''~\cite{rice1996redlining}. These notions were selected due to their legal~\cite{affirmative_anti_discri2011,rice1996redlining,wachter2021cdp}, and practical~\cite{executive2016big,wallstreet2010,toon2012book,barocas_fairmlbook} significance in socio-economic contexts. Throughout the exploration, we use the popular Adult dataset~\cite{adult} as $\mathcal{D}$,  and for each fairness notion employ an in-processing fair-classifier~\cite{agarwal2018} $h(X)$ constrained on respective fairness notion. 

\begin{figure*}
    \centering
    \includegraphics[width=0.9\linewidth]{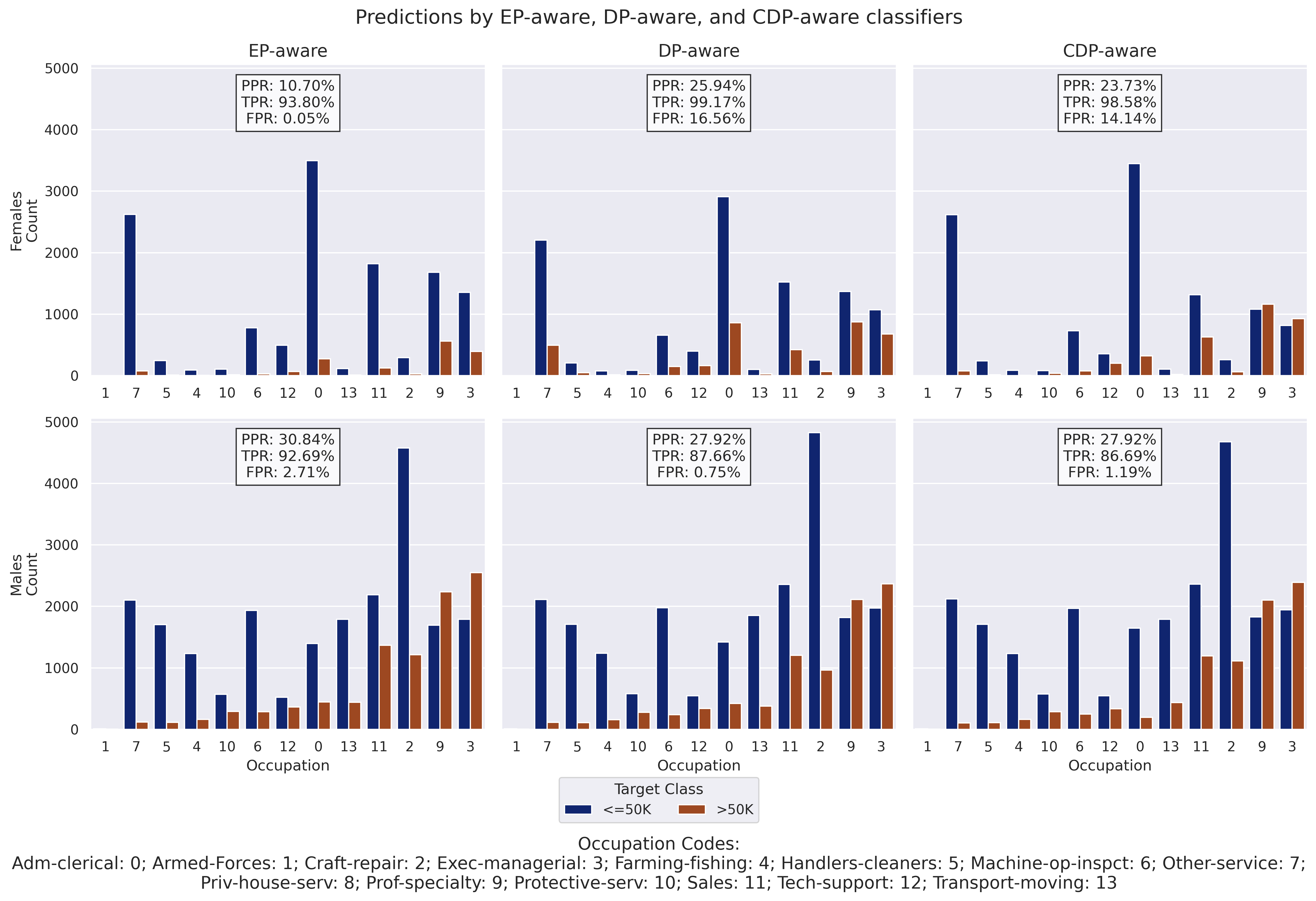}
    \caption{Distribution of positive (>50K) and negative ($\leq$50K) classes for males and females predicted by EP-aware, DP-aware, and CDP-aware classifiers. Occupations are sorted in ascending order w.r.t. count of positive (ground truth) labels from left to right.}
    \label{fig:dist_by_ocu}
\end{figure*}

\begin{figure*}
    \centering
    \includegraphics[width=0.9\linewidth]{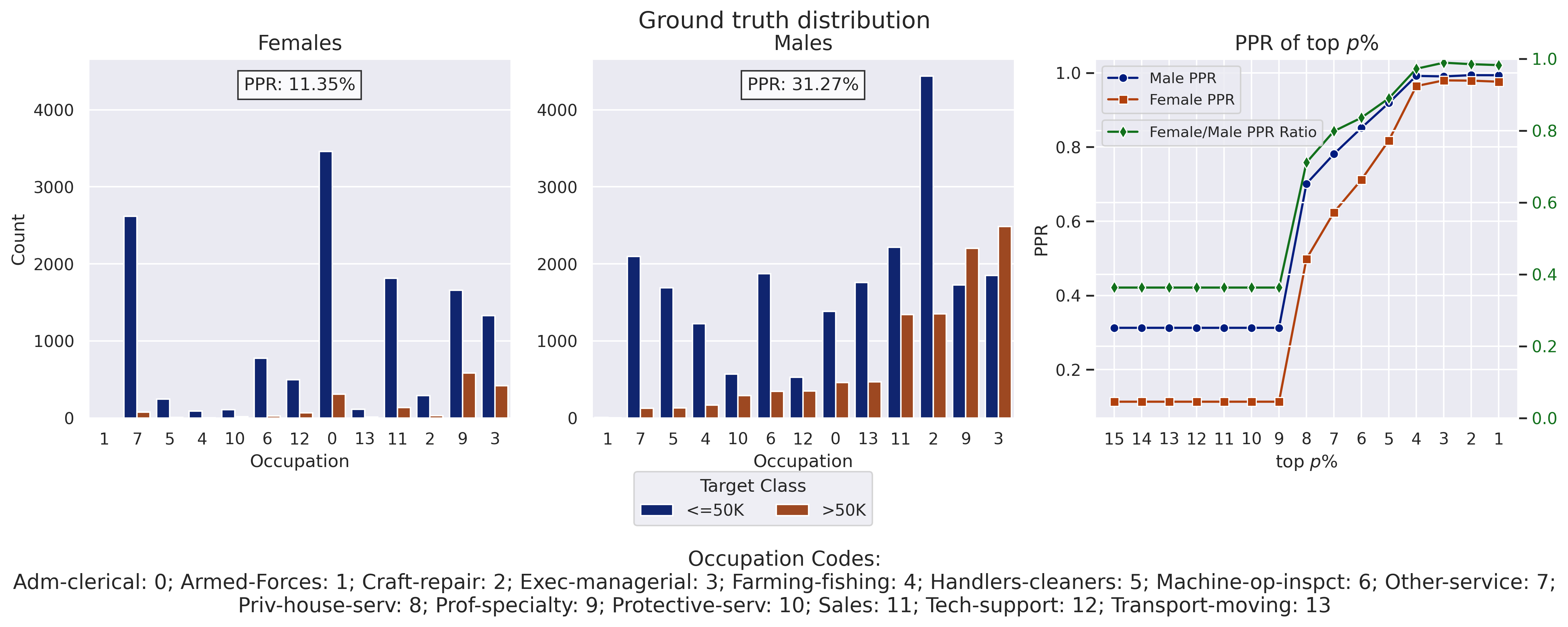}
    \caption{Ground truth distribution of positive (>50K) and negative ($\leq$50K) classes for males and females by occupation. Occupations are sorted in ascending order w.r.t. count of positive labels from left to right. PPR scores refer to the \% of positive labels in the ground truth. PPR of top $p$\% represents the population with top $p$\% capital gain.}
    \label{fig:dist_by_ocu_true}
\end{figure*}

\textit{\textbf{Equal Opportunity (EP)}\cite{hardt2016equality}:}
Eq.\ref{eq:generic_fair} specifies EP under $C = (Y=1)$, i.e., conditioned on the positive ground truth: 
\small$\mathbb{E}(h(x)\mid y=1)=\mathbb{E}(h(x)\mid s,y=1)$\normalsize. 
While EP promotes prediction correctness across demographics, it often struggles with class imbalances in fairness-aware datasets~\cite{tai2022survey}. For example, Fig.\ref{fig:dist_by_ocu} shows that an EP-aware classifier\cite{agarwal2018} can maintain high TPR ($P(h(X)=1\mid Y=1)$) and low FPR ($P(h(X)=1\mid Y=0)$) for both sex yet still exhibit substantial gaps in PPR ($P(h(X)=1)$). Moreover, it exacerbates imbalances against females across \texttt{occupation}, as seen in Fig.\ref{fig:dist_by_ocu}, thereby widening male-female disparities. The ground truth distribution (Fig.\ref{fig:dist_by_ocu_true}) reveals that females experience fewer favorable outcomes (PPR Female/Male$\approx 0.36$) than males. Even with respect to \texttt{capital-gain}—a wealth proxy~\cite{Mitchell2023reportCapitalGain}—females consistently show lower PPR than males with gap being higher for the less wealthier (capital-gain below top 8\%). Thus, EP by emulating ground truth, can amplify entrenched socio-economic inequalities.


\textit{\textbf{Demographic Parity (DP)}~\cite{dwork2012fairness}:}
DP is achieved when \small$\mathbb{E}(h(x))=\mathbb{E}(h(x)\mid s)$\normalsize~ in Eq~\ref{eq:generic_fair}, ensuring outcomes are independent of the protected attribute. Originally motivated by the need to combat racial discrimination in financial contexts (e.g., unequal credit)\cite{wallstreet2010}, DP aims to yield nearly equal PPR across all demographics, as shown by the DP-aware classifier\cite{agarwal2018} in Fig.~\ref{fig:dist_by_ocu} (PPR - females: 25.94\%, males: 27.92\%). 
This often relies on higher false positives for the disadvantaged group, resembling affirmative action'\footnote{“Affirmative action” or “positive action” (as referred to in the EU anti-discrimination Directives) refers to specific measures favouring members of a disadvantaged group~\cite{european2018handbook,affirmative_anti_discri2011}.}, yet its blind distribution of positive decisions raises concerns about exacerbating negative prejudices against protected groups~\cite{barocas_fairmlbook}. As an example in Fig.~\ref{fig:dist_by_ocu}, on high-income (\#positive ($>$50K) $>$ \#negative ($\leq$50K)) professions for males (c.f. Fig~\ref{fig:dist_by_ocu_true}) such as Executive managerial’ and `Professional specialty’, DP-aware approach fails to address class imbalances for females, missing opportunities for more targeted positive action that could help mitigate existing disparities. 


\textit{\textbf{Conditional Demographic Parity (CDP)}~\cite{toon2012book}:} 
CDP extends DP by conditioning equality in PPR on a categorical unprotected attribute (denoted by $X_a$). More formally based on Eq~\ref{eq:generic_fair}:  $\mathbb{E}(h(x)\mid a)=\mathbb{E}(h(x)\mid s,a)$ for all $a \in X_a$. CDP draws inspiration from study of ``redlining''~\cite{rice1996redlining} effect, where location proxies led to financial service disparities. For Adult, we trained a CDP-aware classifier with ``occupation'' as the conditional attribute.
As shown in Fig.~\ref{fig:dist_by_ocu}, it achieves nearly equal PPR for males (0.28) and females (0.24) with high TPRs (>0.86). 
Unlike EP- and DP-aware classifiers, the CDP-aware model achieves fairer outcomes across occupations (Fig.~\ref{fig:dist_by_ocu}), predicting a higher likelihood of high income for females in high-income professions.

\textit{\textbf{Should CDP be the gold standard of fairness in socio-economic contexts?}}
Fig.~\ref{fig:dist_by_ocu} demonstrates that the CDP-aware classifier better promotes protected demographics. However, it remains unclear whether  its decisions are influenced by factors such as SES-privilege. Socio-economic privilege, including inherited wealth~\cite{korom2016inherited,morck2000inherited}, often impacts opportunities~\cite{haneman2021wealth}. Although the Adult dataset lacks some explicit wealth attribute, the \texttt{capital-gain} attribute could serve as a proxy~\cite{Mitchell2023reportCapitalGain}. Studies~\cite{panagiotou2024tabcf} suggest a strong correlation between capital gain and favourable outcomes. 

\begin{figure*}
    \centering
    \includegraphics[width=0.9\linewidth]{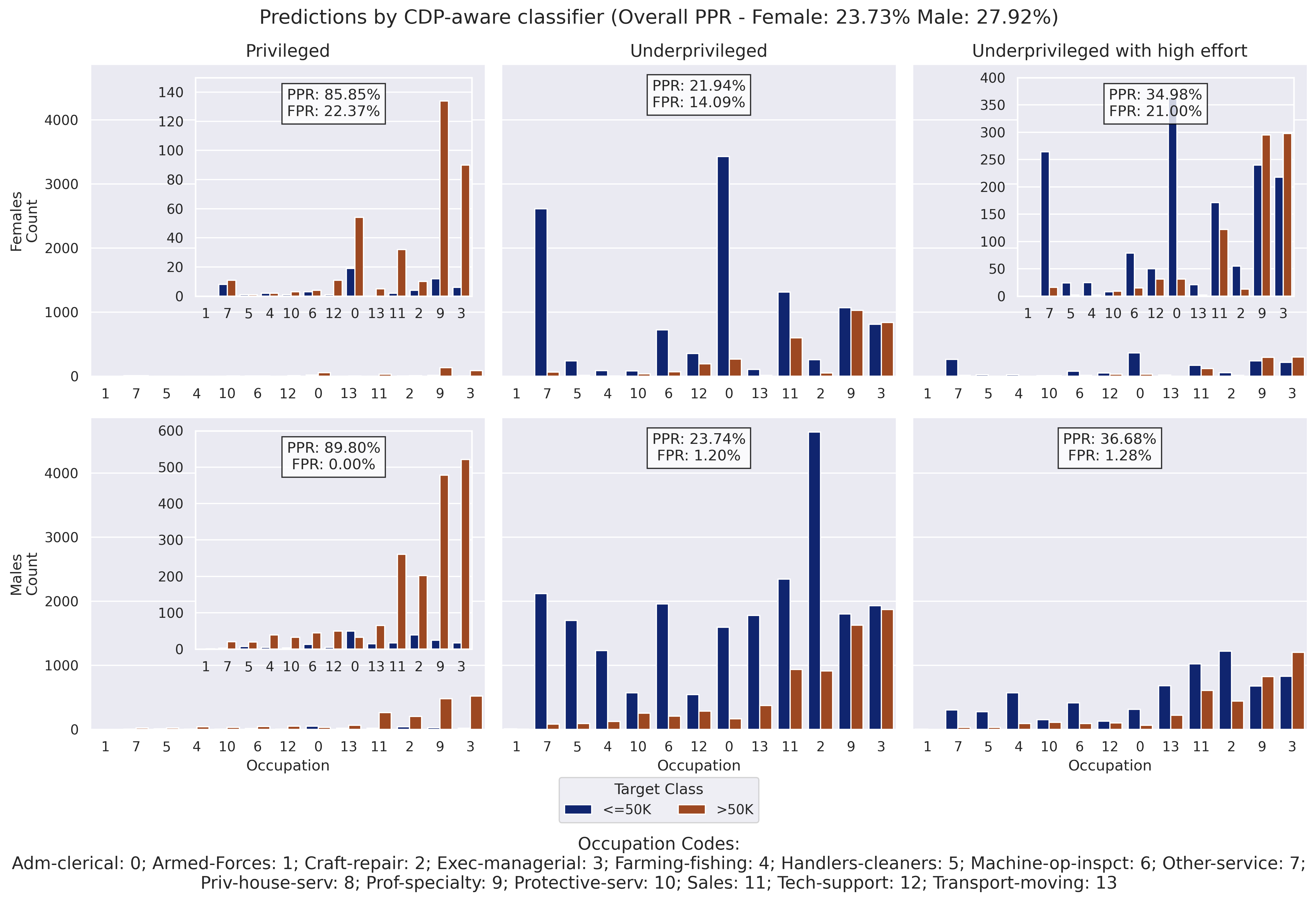}
    \caption{Predictions by CDP-aware classifier on female and male demographics, across different occupations, for socio-economic subgroups defined as privileged (individuals among top 5\% capital gain), underprivileged (individuals below the top 5\% capital gain margin), and underprivileged with high efforts (underprivileged individuals with working hours per week greater than mean of their demographics). Subplots that have fewer data (due to data scarcity of the subgroup) are presented with a zoomed version inside the respective subplots to provide a better view of the results.}
    \label{fig:work_hour}
\end{figure*}

Using capital gain, we define the  ``Privileged'' group as the top 5\% in capital gain and the ``underprivileged'' group as the remaining population. In Fig.~\ref{fig:work_hour}, we observe stark disparities between these groups across males and females. In particular, privileged males and females achieve high PPRs (89.80\% and 85.85\%, respectively), whereas underprivileged groups see sharp declines (23.74\% for males and 21.94\% for females, respectively). Positive actions for underprivileged females (through the CDP-aware classifier) contribute to higher false positives (FPR 14.09\%) compared to underprivileged males (1.20\%), which some may argue is necessary to promote equity. However, as already mentioned, concerns arise about selecting undeserving candidates, potentially increasing biases against protected groups~\cite{barocas_fairmlbook}.

    To address this, we evaluate individual effort using a proxy, the \texttt{hours-per-week} attribute\footnote{Work hours in this example is just used as a reference and not necessarily our view-point. In reality measuring effort~\cite{cappelen2009rewarding} can vary depending on policy. Our SEP and CSEP notions can comply with any such real-valued function measuring effort.} in Adult. In particular, we define individuals above the mean as ``high-effort individuals''~\cite{becker1985effort}. Fig.~\ref{fig:work_hour} shows improved PPR for high-effort subgroups (34.98\% for females and 36.68\% for males, respectively). Still, discrepancies in FPR (\textbf{6\%$\downarrow$}) and PPR (\textbf{59\%$\downarrow$}) between privileged and the underprivileged females show unfair treatment due to the privilege factor.

In conclusion, privileged groups benefit significantly from CDP, while underprivileged individuals, even with high effort, remain disadvantaged. Following the principle of equalization~\cite{dworkin2018equality}, affirmative actions should prioritize effort and need over privilege. 
Although, the CDP-aware classifier, addresses some disparities, it fails to explicitly account for wealth disparities and does not adequately reward high-effort, underprivileged individuals. 
Therefore, CDP may not suffice as the gold standard of fairness in socio-economic contexts. 
Building on these observations, we propose a new fairness notion that aims to capture both (demographic) need and (individual) effort by explicitly integrating SES  and effort into its framework.

\section{SOCIO-ECONOMIC STATUS-AWARE FAIRNESS NOTION}\label{sec.method}
Our findings in Sec. \ref{sec.background}  suggest that to be fair in the socio-economic context certain social factors, such as inherited privilege, and economic factors such as individual effort, need to be considered. More concretely:
\begin{itemize}
    \item[a.] The chances of favourable outcomes should not be hindered for socio-economically underprivileged individuals.  
    \item[b.] Individuals in underprivileged subgroups who put higher effort beyond a threshold should be given a higher chance of positive rewards, ideally  proportional to their effort.
    \item[c.] Positive action for privileged subgroups should be minimized in order to reduce the influence of privilege.
\end{itemize}
Building on these observations, we define two notions of SES-aware fairness (Socio-Ecomomic Parity - SEP and its stronger counterpart, Conditional SEP - CSEP) (Sec.~\ref{sec.def_sep}) and their corresponding  
SES-aware fairness metrics (Sec.~\ref{sec.measure_sep}). We train fair models based on these metrics and evaluate their effectiveness in mitigating discrimination (Sec.~\ref{sec.our_emp}). 

\subsection{\textbf{Defining Socio-Economic Parity (SEP)} and Conditional Socio-Economic Parity (CSEP)}
\label{sec.def_sep}

\begin{definition}
  A predictor $h$ is said to behold Socio-Economic Parity (SEP) fairness if for any given economically privileged group defined as $(X_p\geq \tau_p)$, and underprivileged sub-groups defined as $(X_p< \tau_p,S=s)$ by a socio-economic attribute $X_p$, a privilege threshold $\tau_p$, a protected attribute $S$, and demographic identity $s$, it holds: 
\begin{equation}\label{eq.sef}
\begin{split} 
    &\mathbb{E}(h(x)) = \mathbb{E}(h(x)|s, x_p<\tau_{p}), \text{\phantom{space}}
    \forall {(x,s,y)}\in D; ~~\zeta: \mathbb{R} \to [1,\infty) \\
    &\text{such that:\phantom{to:} }\\ 
    &\text{\phantom{to:}}
    \mathbb{E}(h(x)\cdot \zeta(x_e)\mid x_p<\tau_{p}, x_e\geq e)\geq \mathbb{E}(h(x)\mid x_p<\tau_{p}, x_e< e),\\
    &\text{\phantom{to:}} \mathbb{E}(h(x)\mid x_p\geq\tau_{p},y=0)\leq \mathbb{E}(h(x)\cdot \zeta(x_e)\mid x_p<\tau_{p}, x_e\geq e,y=0)
\end{split}
\end{equation}\normalsize
\end{definition}
where $x_e$ is the value w.r.t. $X_e$-an identified non-protected attribute that can be expressed as a proxy for measuring effort put forth by an individual, $\epsilon$ is the disparity tolerance (ideally 0), and $\zeta{(\cdot)}$ is a real-valued function that relatively weights underprivileged individuals (to $\geq$1) based on their effort value $x_e$ for individuals with $X_e\geq e$. The threshold $\tau_{p}$ defines the privileged group as the population with top $p\%$ value distribution of $X_p$. 

In simpler words, SEP is achieved when the expected favourable outcome for any protected group $s$ is independent of socio-economic privileges. Additionally, it ensures that the false positive rate for the privileged group is not higher than that of the underprivileged group exerting high effort. Furthermore, the cumulative rewards for underprivileged high effort individuals are proportionally weighted based on their evaluated effort.   
%
%
%
A stronger (and more desirable) version of the SEP fair notion, is what we call Conditional Socio-Economic Parity (CSEP):
\begin{definition}
    A predictor $h$ thats behold Socio-Economic Parity (SEP) across all the categories $a$ defined by a categorical non-protected attribute $X_a: X_a\neq X_p,X_e$, is called Conditional Socio-Economic Parity (CSEP) -fair. Formally:
\begin{equation}\label{eq.csep}
\begin{split}
    &\mathbb{E}(h(x)\mid a) = \mathbb{E}(h(x)|s, a, x_p<\tau_{p}),\\ &\text{\phantom{such that}}\forall_{a\in X_a, X_a\neq X_p,X_e}, \forall {(x,s,y)}\in D ; ~~\zeta: \mathbb{R} \to [1,\infty)\\ 
    &\text{such that: \phantom{to:}  }
    \\ 
    &\text{\phantom{}}
    \mathbb{E}(h(x)\cdot \zeta(x_e,a)\mid a, x_p<\tau_{p}, x_e\geq e)\geq \mathbb{E}(h(x)\mid a, x_p<\tau_{p}, x_e< e),\\
    &\text{\phantom{}} \mathbb{E}(h(x)\mid a, x_p\geq\tau_{p},y=0)< \mathbb{E}(h(x)\cdot \zeta(x_e,s,a)\mid
    \\&\text{\phantom{space, space}}a, x_p<\tau_{p}, x_e\geq e,y=0)
\end{split}
\end{equation}\normalsize
\end{definition}
Note that in Eq.~\ref{eq.csep}, the function $\zeta(\cdot)$ takes the additional argument $a$. This is to ensure that the effort-based weighting also considers the localized condition. 
For example, in each job, effort weights will vary based on the threshold, such as the average working hours per week in that job. 
The notion can be further refined by adopting subgroup-specific thresholds for $\tau_p$ and $e$, in order to more accurately capture the socio-economic influences within each community.

\subsection{Measuring the violation of SEP and CSEP}\label{sec.measure_sep}
Based on definitions of the SEP, and CSEP notions introduced in Sec.~\ref{sec.def_sep}, here we introduce the AI fairness metric measuring the SES-aware bias/discrimination. Following Eq.~\ref{eq.sef}, violation of SEP can be measured as:
\small
\begin{equation}\label{eq.m_sep}
    \begin{split}
      \forall&{s\in S},~~~~~ |P(h(X)=1) - P(h(X)=1\mid X_p<\tau_p,S=s)|\\+ &|\frac{1}{A}\sum_{x: x_e<e }P(h(X=x)=0\mid X_p<\tau_p, S=s, X_e < e)\\- &\frac{1}{B}\sum_{x:x_e\geq e}\zeta(e,s)P(h(X=x)=0\mid X_p<\tau_p, S=s, X_e\geq e)|\\
        + &|\frac{1}{C}\sum_{x:x_p\geq \tau_p}P(h(X=x)=0\mid X_p\geq\tau_p, Y=0)\\ - &\frac{1}{B}\sum_{x: x_e\geq e }\zeta(e,s)P(h(X=x)=0\mid X_p<\tau_p,s, X_e\geq e, Y=0)| \leq \epsilon
   \end{split}
\end{equation}
\normalsize
where \small$A$=$\sum\limits_{x_p< \tau_p,x_e<e}{\mathbf{1}}$, $B$=$\sum\limits_{x_p< \tau_p,x_e\geq e}{\zeta(e,s)}$,\normalsize~and \small$C$=$\sum\limits_{x_p\geq \tau_p}{\mathbf{1}}$,\normalsize~ensures the respective parts remains a probability distribution. 
Note that the second and the third part of the Eq.~\ref{eq.m_sep} focuses on equating the probability negative predictions i.e., $h(X=x)=0$, contrary to the favourable (positive) outcome expectation goal in Eq.~\ref{eq.sef}. This is intentional and necessary, to avoid penalizing individuals exerting high effort. 
Similar to Eq.~\ref{eq.m_sep} we can trivially extend the measure following Eq.~\ref{eq.csep} to evaluate CSEP violation conditioned on non-protected condition $X_a$. 

In absence of mechanism to evaluate effort, or to simplify computation and interpretation, violation of SEP (Eq.~\ref{eq.m_sep}) can be relaxed to:
\small $  \forall{s\in S},~~~~  |P(h(X)=1) - P(h(X)=1\mid X_p<\tau_p,s)|\leq \epsilon$
\normalsize
where the relaxed goal is to only capture the discrimination towards all the underprivileged demographic subgroups. 
\begin{figure*}
    \centering
    \includegraphics[width=0.9\linewidth]{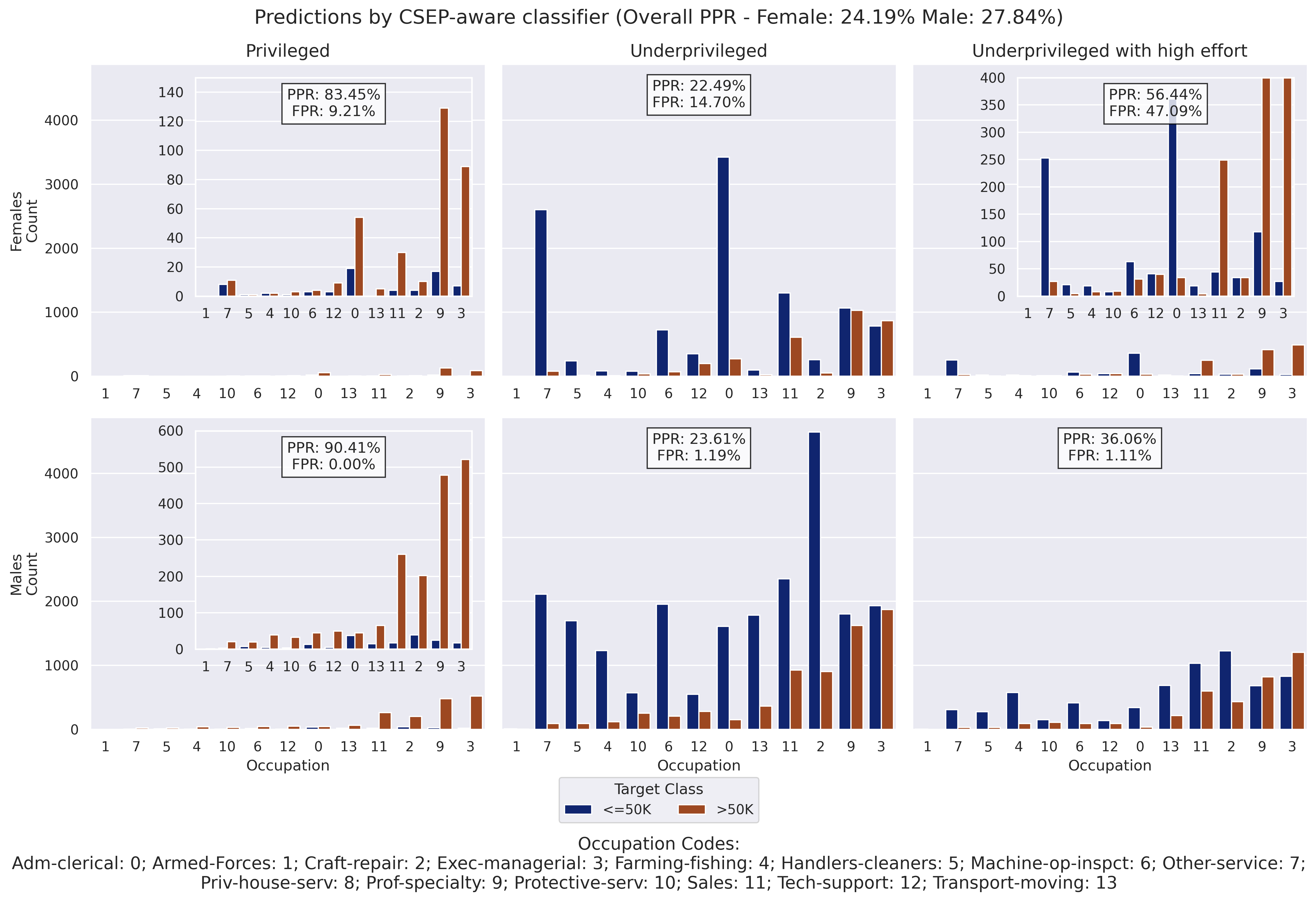}
    \caption{Distribution of positive (>50K) and negative ($\leq$50K) classes for males and females predicted by CSEP-aware classifier. Occupations are sorted in ascending order w.r.t. count of positive (ground truth) labels from left to right.}
    \label{fig.our_on_adult}
\end{figure*}
\subsection{Empirical  effectiveness of SES-aware fairness}\label{sec.our_emp}
\begin{figure*}
    \centering
    \includegraphics[width=\linewidth]{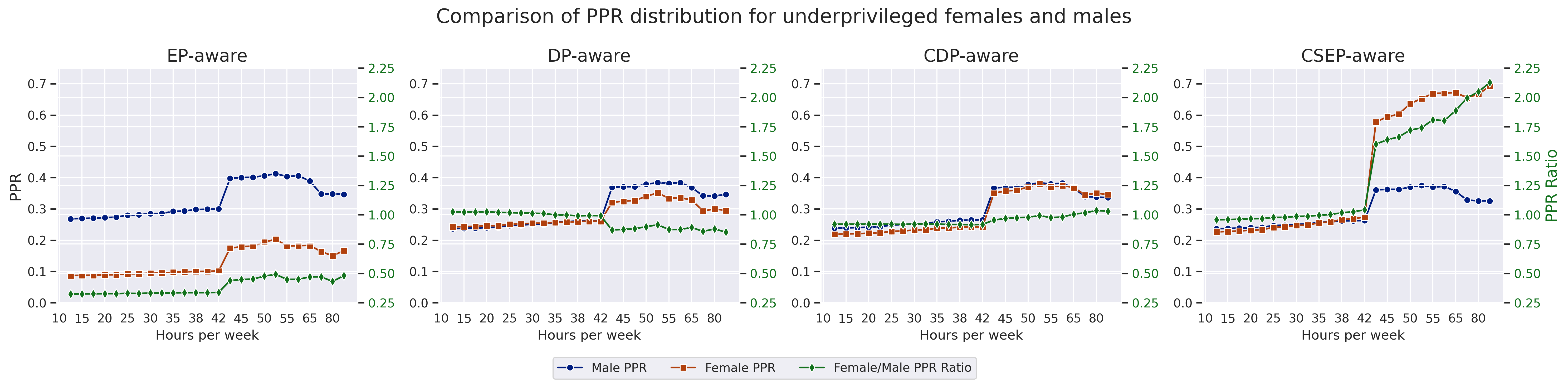}
    \caption{Comparison of our CSEP-aware model against EP, DP, and CDP -aware models in distributing positive action towards  underprivileged females and males. Higher PPR ratio indicates higher preference of positive action for females.}
    \label{fig.positive_action_comparison}
\end{figure*}

To examine the effectiveness of our CSEP notion in reducing the effect of SES, similar to our previous exploration, we optimize a model~\cite{agarwal2018} constrained on our notion (Eq.~\ref{eq.csep}) and test it on the Adult data. We plot the results in Fig~\ref{fig.our_on_adult}. The key things to notice here is that our CSEP-aware model starkly improves the PPR of high-effort underprivileged protected group to 56.44\% (by 59\%$\uparrow$) by awarding more positive actions for this niche subgroup (FPR 47.09\%) while keeping the false predictions for the overall underprivileged females at par with that of CDP-aware model (Fig.~\ref{fig:work_hour}). Further, our CSEP-aware model also brings down the positive actions for the privileged  females (FPR 58\%$\downarrow$) while keeping the overall PPR between male (27.84\%) and female (24.19\%) nearly equal.  
Fig.~\ref{fig.positive_action_comparison} further compares PPR rates by hours per week for underprivileged males and females under EP-, DP-, CDP-, and CSEP-aware models. Notably, CSEP allocates positive outcomes proportionally to effort among high-effort underprivileged females ($\geq$ mean effort), nearly doubling the female-to-male PPR ratio (0.98 to 2.1) and mitigating both overall demographic disparities and those observed among protected underprivileged subgroups (Fig.~\ref{fig:dist_by_ocu_true}). Specifically, the PPR for underprivileged females increases from below 0.3 for those working fewer than the mean hours to above 0.6 for those exceeding it, reaching nearly 0.7 for those working 55 hours\footnote{Work hours in this example is just used as a reference and not necessarily our view-point. In reality measuring effort~\cite{cappelen2009rewarding} can vary depending on policy. Our SEP and CSEP notions can comply with any such real-valued function measuring effort.} or more. In contrast, EP- and DP-based methods under-reward these high-effort females, resulting in lower PPRs than their male counterparts. Although the CDP-aware model provides higher PPR for all underprivileged individuals, it does so uniformly across both demographics, failing to deliver the additional positive actions needed to support high-effort underprivileged females and address remaining disparities. 
\section{INVESTIGATING SEP NOTION UNDER THE AI ACT AND EU ANTI-DISCRIMINATION LAW 
}\label{sec.features that define}
By placing the provision of justifications at the forefront of the study, we aim to ensure accountability, as the definition and realisation of explicit selections of the important attributes, context and any trade-offs of a fairness-aware metric undertaken to protect the rights, freedoms, and interests of individuals \cite{information_commissioners_office_ico_guidance_2024}. Thus, drawing upon the elucidation of the field of SES under the AI Act, the regulation states that ``...\textit{AI systems used to evaluate the credit score or creditworthiness of natural persons should be classified as high-risk AI systems}...'' (Recital 58). 
In parallel, from the combination of Article 6(2) and Annex II, par. 5(b) of the AI Act, it is explicitly indicated that AI systems ``\textit{intended to be used to evaluate the creditworthiness of natural persons or establish their credit score, with the exception of AI systems used for the purpose of detecting financial fraud}'' are classified as high-risk AI systems, with the requirement to satisfy the provisions laid down in Articles 8-15 (Chapter III, Section 2) that should be assessed under the conduction of a conformity assessment [Article 3(20)]. 
In addition to those obligations, deployers of the systems that are described in Article 6(2) should perform a Fundamental Rights Impact Assessment (FRIA) and notify the national authority of the results (under the circumstances of Article 27); thus, regarding AI bias, the FRIA aims to detect and mitigate the discriminatory risks of an AI system.

Regarding the topic of AI bias in AI systems, the AI Act formalises the requirement of identification and mitigation of unfair discrimination in high-risk AI systems [Article 10(2)(f), (g)]. Especially for the aspect of detection, AI system providers should first identify the risk level of each AI system, recognising also the importance of preventing the (re)production of biases, that can lead to discrimination and taking into consideration the potential protected ground(s) of discrimination under EU law; the requirement of representative and ad hoc datasets [Article 10(3),(4)] also contributes to that direction. Moreover, the requirement for human oversight in Article 14(4)(b) of the AI Act shifts the focus to the identification of automation bias by the responsible person ``in the loop'' in the context of high-risk AI systems. Regarding accuracy, robustness and cybersecurity, Article 15(4) introduces the requirement to apply bias mitigation measures to the development of high-risk AI systems in order to avoid reproducing bias via ``feedback loops''. Furthermore, accountability is enhanced through the requirements of the prior design of the technical documentation (Article 11) and record-keeping of logs (Article 12). 

\begin{figure*}
    \centering
    \includegraphics[width=0.8\linewidth]{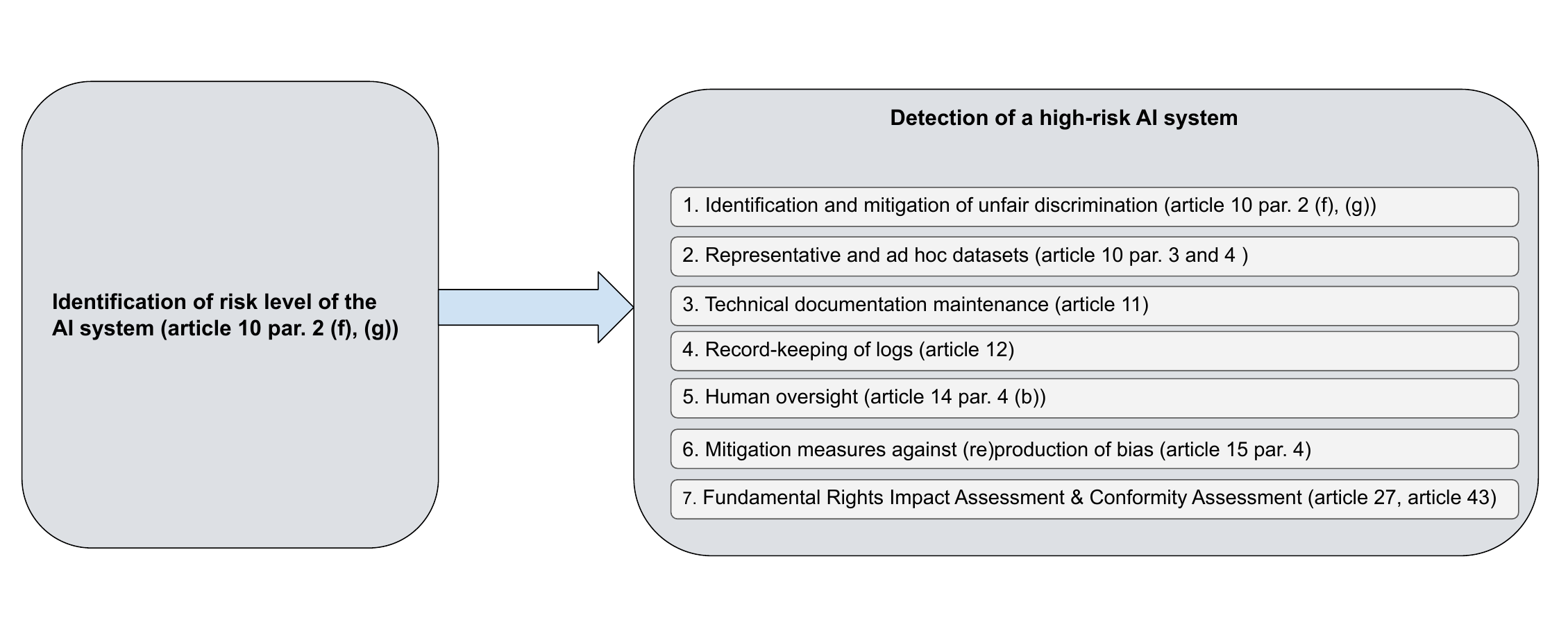}
    \caption{Specific provisions in the AI Act addressing AI bias}
    \label{fig:bias provisions}
\end{figure*}

Therefore, AI fairness-aware metrics should be designed through the notions of proving compatibility with the fundamental rights in order to be able to be utilized through the required assessments of the AI Act. The right not to be discriminated, which is a fundamental principle enshrined in EU law, should be explored and examined by every attempt to address AI bias. From the text of the AI Act, it is obvious that the regulation delegates the determination of discrimination to existing EU legislation \cite{deck2024implications}.

The two main forms of discrimination, under EU law, include direct and indirect discrimination \cite{zuiderveen2024non,wachter2020bias}. Direct discrimination\footnote{Relevant provisions to EU law are included in Article 2(2)(a) of the Directive 2000/43/EC, Article 2(2)(a) of the Directive 2000/78/EC, Article 2(1)(a) of the Directive 2006/54/EC and Article 2(a) of the Directive 2004/113/EC, to the extent of the protected grounds of the directives.} occurs when a person is being undertaken less favourable treatment than another in a comparable situation, based on a protected ground, while indirect discrimination\footnote{Relevant provisions to EU law are included in Article 2(2)(b) of the Directive 2000/43/EC, Article 2(2)(b) of the Directive 2000/78/EC, Article 2(1)(b) of the Directive 2006/54/EC and Article 2(b) of the Directive 2004/113/EC, to the extent of the protected grounds in the directives.} exists when an ostensibly neutral provision, criterion or practice sets people with a particular protected characteristic in a disadvantaged position compared with others unless that provision, criterion or practice is objectively justified by a legitimate aim and the means of achieving that aim are appropriate and necessary\footnote{Article 2(2)(b) Directive 2000/43/EC; Article 2(2)(b) Directive 2000/78/EC; Article 2(b) Directive 2004/113/EC; Article 2(1)(b) Directive 2006/54/EC.} \cite{noauthor_bias_2022,european2018handbook}. Therefore, establishing indirect discrimination requires demonstrating the presence of two groups (comparators), one advantaged and one disadvantaged by the contested measure \cite{european2018handbook}. An expression of direct discrimination is discrimination by association \cite{gerards2021algorithmic}, which can be identified when a person is treated less favourably due to another person’s protected characteristic\footnote{See: CJEU, C-303/06, S. Coleman v. Attridge Law and Steve Law, 17 July 2008.}. In cases where several grounds of discrimination exist, multiple discrimination takes place when the grounds operate separately, and intersectional discrimination occurs when the grounds interact, are inseparable and produce specific types of discrimination; nevertheless, the Court of Justice of the European Union (CJEU) has not recognized that a new category of discrimination can arise from the combination of more than one of those grounds \cite{european2018handbook,RoyFact23}.

It has been widely claimed in the literature that AI bias should be viewed and addressed through the concept of indirect discrimination \cite{zuiderveen2020strengthening,barocas2016big, zarsky2017analytic}. With the reasoning that direct discrimination will arise only if the explicit or implicit bias of the decision-maker influences the model, cases of unintentional discrimination, such as those caused by sampling errors or historical bias, are excluded from consideration \cite{hacker2018teaching} and due to the nature of AI systems, as they depend on inferences and proxies for target variables and protected attributes \cite{wachter2020bias}.

\subsection{\textbf{Automated contextual extraction of source of privilege ($X_p$)}}

Starting from the principle of proportionality under indirect discrimination, once it has been established, the burden of proof shifts to the defendant, effectively initiating a proportionality test \cite{gerards2021algorithmic}. It has been argued that AI fairness metrics should be based on the ``proportionality test'' (where the legitimate interest is pursued in a manner that is both necessary and proportionate) \cite{wachter2020bias}; this approach can enhance the alignment between legal principles and AI-fairness metrics and, per analogiam, in the determination of the trade-offs and essential balancing of interests through fairness-aware mechanisms. Therefore, our solution includes the implications and essential justifications for the following question:

\textit{\textbf{Are the selected attributes of the AI fairness-aware approach/ metric necessary and appropriate to protect the legitimate interests of the protected groups}?}\footnote{``necessary'' and ``appropriate'' means that they ``...must be suitable to achieve the desired end'' \cite{propotionality}.} 
From a top-down approach, we start by identifying the source of biases that can affect the protected groups. The attributes identified by our approach, as most critical for accurately representing the comparators’ groups, depend on statistical data and the ability of the trained models to identify bias. By implementing computational fairness notions to compare the treatment of different groups, clear statistical evidence of the existence of discrimination can be provided; this evidence can be taken into consideration by the court in cases involving allegations of discrimination\footnote{See Recital 15 of the Racial Equality Directive 2000/43/EC.} \cite{zuiderveen2024non,fredman2011discrimination}.

Also, it is revealed that the attributes (including protected grounds and proxies) that are pivotal to influence the decisions as most critical for accurately representing the comparators’ groups and prone to affect the model's learning bias, as the contribution of determining proxy variables, which are non-protected grounds (‘capital-gain’, ‘work-hour-per-week’, ‘occupation’) can guide the AI fairness-aware measures, through the investigated correlations to acknowledge protected grounds (‘sex’) and, as a result, contribute to the aim of definition and transforming the harmful discriminatory effects in SES context. The fact that the ground of SES is not covered by Article 19 TFEU \cite{ganty2021expanding} has provoked discussions on multiple levels; primarily, the lack of SES recognition \cite{ganty2021expanding}, and secondarily, in algorithmic discrimination through harmful patterns that are present and do not directly relate to protected grounds under EU law \cite{eubanks2018automating}.

Identifying attributes that can be considered as a source of privilege requires background knowledge about what can constitute socio-economic privilege and what is the historical impact of the considered attribute in determining the current outcome. So, 
some may rightly argue that a fully automated determination of $X_p$ may not be possible. However, since here we are interested in fairness-aware learning of an AI model, it can be intuitively useful if we focus on what the AI model considers as source of privilege. One simple strategy would be to identify an unprotected attribute which is important in producing favourable outcome for a non-protected demography (historically known to garner privilege), and inherently with its attribute values can segregate the population into different sections of hierarchy. We use a simple two-step process to approximate the above-mentioned idea, given as follows:
\begin{itemize}
    \item[\textbf{step1:}] Train a predictor $h_g(\cdot)$ with a subset $\mathcal{D}_g$ i.e., the subset of the data $\mathcal{D}$ drawn only from the distribution of the non-protected population $\mathcal{P}(X,g,Y)$. 
    \item[\textbf{step2:}] Among the subset of unprotected  ordinal ($X_o$) e.g., ``Education''~\cite{arbel2017equaleducation}, and numerical ($X_n$) e.g., ``Capital gain''~\cite{williams2017wealth} attributes, using a feature importance method $\mathcal{FI}$~\cite{ewald2024FI}, identify the attribute $X_p\in (X_o \bigcup X_n)$ that has the highest impact on model's predictions for the privileged population $\mathcal{D}_g$, i.e., \small$X_p$=$\argmax\limits_{X_i\in(X_o\bigcup X_n)}\mathcal{FI}(h(X),X_i)$\normalsize. Note that the restriction of attribute $X_p$ being either ordinal or nominal is a necessary condition to quantify the hierarchy of privilege.  

\end{itemize}

\subsection{Determining the value of $p$, which top $p$\% privileged population to consider?}

The identification of the comparators’ groups plays a vital role both in the legal justification of the discrimination and in the implementation of the proposed fairness-aware notions. First, our approach initiates a comprehensive statistical investigation which is used to reveal the discrimination patterns and emphasizes the context-specific implementation of the fairness transforming methods, aiming to satisfy the goal of addressing AI bias with the ad hoc appropriate actions, which are considered proportionate. To align with the definition of indirect discrimination and address it in practice, we classify individuals into two comparator groups (``privileged'' and ``underprivileged''), based on the criterion of disadvantaged SES. Therefore, the two groups, based on sex, have been determined according to the variables of \texttt{capital-gain}, \texttt{work-hour-per-week}, \texttt{occupation}. Second, our investigation shows in Fig. \ref{fig.positive_action_comparison} that if the \texttt{underprivileged group}, when given certain advantages as the privileged, can perform equally based on the related variables. To enable this prediction, we set other two subgroups indicating the top $p$\% in relation to the advantage of high capital gain. The positive prediction is based on counterfactual rationale \cite{byrne2019counterfactuals}, where these females will have a higher capital gain (if we increase the attribute value for the identified females that do not belong to the top $p$\%). It should be mentioned that this counterfactual reasoning lays down a ``positive action''\footnote{Relevant provisions: Article 5 Directive 2000/43/EC; Article 7 Directive 2000/78/EC; Article 6 Directive 2004/113/EC; Article 3 Directive 2006/54/EC.} which refers to the measure that promotes preferential treatment for disadvantaged groups \cite{o2011positive} to eliminate, prevent or remedy past discrimination \cite{eurofound_positive_2011}. As ``positive action'' is, in general, adopted under EU law, the requirement should be for the AI fairness metrics to include the balancing to reach the goal of equal treatment, by not maintaining feedback loops that can deteriorate the SES of a population group. 

Third, the determination of the comparators’ groups based on a threshold (top 5\% of males and females), in the case of fairness-aware methods, relies on a quantitative criterion. Initially, it should be mentioned that the evidential basis of the criterion aligns with this requirement of CJEU\footnote{CJEU, C-167/97, Seymour-Smith, 9 February 1999.} with the aim of displaying that the contested rule is related to any form of discrimination \cite{hacker2018teaching}. Apart from clarifying the measurable nature of the predictive actions, the focus should also be shifted to its examination of satisfying the characteristics of ``necessary'' and ``appropriate''. In general, for the determination of comparators' groups through the definition of a threshold, it has been claimed that explicit quantitative limits derived from the EU legal framework are absent \cite{sandra2011discrimination}, and a hypothetical disadvantage is sufficient \cite{hacker2018teaching}. The necessity of determining this threshold lies in the objective of identifying the privileged category within the entire population, irrespective of protected identity. In particular, the positive class ratio between females and males surpasses the 80\% rule only from the top $p$\% onward. In parallel, the positive label distribution for the group of females exceeds the 80\% rule, thereby qualifying to be considered a privileged category, from the threshold of the top $p$\% (as it is shown in Fig. \ref{fig:dist_by_ocu_true}), maintaining at the same time the appropriate size of the population. 

\section{STANDARDIZATION FOR THE AI ACT AND FAIRNESS MEASURES UTILIZATION}\label{sec.standartization}
Initially, with regard to AI standards within the scope of the AI Act, the European Committee for Standardization (CEN) and the European Committee for Electrotechnical Standardization (CENELEC) are considered permanent members of the advisory forum for the regulation, as specified in Article 67 of the AI Act. As the practical application of the AI Act will rely on the development and implementation of harmonized standards \cite{CEN-CENELEC2022}, in response to request \cite{EuropeanCommission2023} by the European Commission, CEN and CENELEC accepted to prepare a work programme of European standards and European standardization deliverables \cite{TaskGroup2024}. Notably, AI conformity assessment and risk management are included, as one of the targets (among others) of Task Group Inclusiveness, while the FRIA is not referred in the non-binding first newsletter \cite{TaskGroup2024}. As part of broader efforts to support standardization for legally and ethically fair AI systems, the critical actions necessitate a dialogue between the fairness-aware metrics and principles in light of the FRIA, and the conformity assessment (as described in Sec. \ref{sec.features that define}). In parallel with the provisions of the AI Act, standardization efforts should also consider the 2019 Ethics guidelines for trustworthy AI \cite{ai2019high} along with the Assessment List for Trustworthy AI \cite{eu_altai_2020}, as the Guidelines are recalled in Recital 27 of the AI Act into the risk-based approach of the regulation.

\textit{\textbf{Interference between fairness-aware machine learning measures and the assessments provided by the AI Act}:}
During the conformity assessment process, several requirements must be examined, as illustrated in Fig. \ref{fig:bias provisions}. In particular, there is a strong interference between the technical measures for ensuring AI fairness and the identification and mitigation of unfair discrimination in high-risk AI systems [Article 10(2)(f),(g)] in order to achieve the aims of the regulation. The detection of AI bias can be identified through technical measures that examine AI fairness, which also according to Article 9(5) of the AI Act can guide the design and development of the AI system to prevent or, where prevention is not possible, mitigate relevant risks \cite{mantelero2024fundamental}. The accountability of AI system providers should be placed at the forefront, supported by essential documentation, justifications regarding the process, and the technical solutions that must accompany this assessment to demonstrate practical compliance.

With regards to the FRIA, the risk assessment requirements [of Article 27(1)(a)-(f)] could play a vital role in technical metrics that address bias in the context of high-risk systems, as their implementation can help to evaluate, justify and measure under subpar. (d) ``\textit{the specific risks of harm likely to have an impact on the categories of natural persons or groups of persons}...'' under statistical measurable criteria. As the FRIAs are expected to be implemented as context-based assessments \cite{mantelero2024fundamental}, we would like to highlight the relevance of this condition, which can thus integrate specialized AI fairness metrics designed for individual contexts into the determination of explicit measures and attributes, as well as the setting of acceptable thresholds. More specifically, in terms of identifying/measuring discriminatory risks under the assessment, the establishment of criteria based on the AI fairness metrics can contribute to evaluating the discriminatory risks that AI bias may pose to individuals. In particular, this can be achieved through the analysis of proxies, enabling the exploration of protected grounds and examination of important correlations that lead to discriminatory risks. Moreover, an additional element of the FRIA that could ultimately involve AI fairness metrics and solutions refers to Article 27(1)(f) and especially on ``\textit{the measures to be taken in the case of the materialization of those risks, including the arrangements for internal governance and complaint mechanisms}.'' The implementation of technical solutions (in this case of the proposed method, concerning SES) could serve as a means to address the issue by reconfiguring AI bias in this area, and thus mitigate risks (under the FRIA) when an AI system exhibits risks related to the presence of discrimination. In parallel, the integration of specific AI fairness technical algorithmic solutions can contribute to enhancing a comprehensive management plan aimed at ensuring compliance, during the phase of addressing and responding to the potentially discriminatory outcomes derived from this assessment.

\section{CONCLUSIONS, LIMITATIONS, AND FUTURE DIRECTIONS}\label{sec.conclude}
In this work, we introduced Socio-Economic Parity (SEP), a novel AI fairness notion, alongside Conditional Socio-Economic Parity (CSEP), to address SES-driven disparities in AI systems, 
under the prism of the EU AI Act. This interdisciplinary study aims to bridge legal science and ML and highlight aspects that require further attention and exploration. 
Our empirical evaluation using Adult demonstrates that CSEP effectively rewards high-effort underprivileged protected groups  while maintaining comparable error rates across the broader underprivileged female population, and accounting for equity across the conditional attribute categories. 
Despite these promising results, further investigation is required to address remaining limitations and ensure real-world applicability:
\begin{itemize}
    \item \textit{Scope of evaluation}
The empirical evaluation is currently limited to Adult, a commonly used benchmark in fairness research. 
Future work should test the proposed framework on additional datasets representing diverse fairness scenarios to ensure its generalizability across different contexts.
\item \textit{Context-specificity}
The proposed framework relies on context-specific definitions and proxies for socio-economic status (e.g., capital gain) and individual effort (e.g., working hours). These  were chosen to illustrate the underlying concepts, however, such definitions and proxies should generally be tailored to the specific problem or domain.
\item \textit{Computational overhead} Additional constraints introduced in the notion increases the training computation, and might need relaxation of tolerance threshold $\epsilon$ for convergence. 
\end{itemize}

Despite these limitations, the framework offers a strong \textit{\textbf{foundation for addressing financial exclusion through the exploration of (un)privileged groups}}. The critical balance achieved by SEP relies on measurable evidence provided by the model, enhancing transparency and supported by comprehensive legal explanations for all decisions. In particular, the introduction of comparators’ groups through the concept of indirect discrimination provides an effective tool for identifying and addressing inequalities between privileged and underprivileged groups within the SES. Empirically, our experiments show that applying an SEP-driven fairness constraint can reward high-effort underprivileged subgroups while maintaining comparable error rates across the broader population, underscoring SEP’s potential to narrow socioeconomic gaps in machine-learning applications without sacrificing performance.

Through our legal analysis, we also emphasized the importance of \textit{\textbf{strengthening SES within the EU AI Act}}. The rapid development of the AI sector, highlights the need to refine SES-related evaluations, given its current status as a non-protected ground under EU law. Enhanced SES protection can be reinforced through AI Act assessments, especially considering discrimination risks. 

A vital next step involves leveraging statistical data from sociological and economic studies focused on SES-specific contexts, which can further solidify SES-aware decisions in AI frameworks. 
We hope this work will inspire further studies on the role of SES in fairness-aware machine learning. 




\section*{ACKNOWLEDGMENTS}
This research work was funded by the European Union under the Horizon Europe MAMMOth project, Grant Agreement ID: 101070285. UK participant in Horizon Europe Project MAMMOth is supported by UKRI grant number 10041914 (Trilateral Research LTD). The work of Arjun Roy is also supported by the EU Horizon Europe project STELAR, Grant Agreement ID: 101070122.





\bibliographystyle{ACM-Reference-Format}
\bibliography{sample}

\end{document}